\documentclass[final,twocolumn,pra,showpacs,preprintnumbers]{revtex4}
\usepackage{bm}
\usepackage{amsmath,amssymb,epsfig,color}

\newcommand{\mathd}{\mathrm{d}}

\newcommand{\tmmathbf}[1]{\ensuremath{\boldsymbol{#1}}}

\newcommand{\tmtextit}[1]{{\itshape{#1}}}

\newcommand{\tmop}[1]{\ensuremath{\text{#1}}}
\newcommand{\bignone}{}
\newcommand{\mathbbm}{\mathbb}

\begin{document}

\title{Comment on ``Quantum linear Boltzmann equation with finite intercollision time''}

\author{Klaus Hornberger}
\affiliation{Max Planck Institute for the Physics of Complex Systems, N{\"o}thnitzer Stra{\ss}e 38, 01187 Dresden, Germany }
\author{Bassano Vacchini}
\affiliation{Universit{\`a} degli Studi di Milano, Dipartimento di Fisica, Via Celoria 16, I-20133 Milano, Italy }
\affiliation{INFN Sezione di Milano, Via Celoria 16, I-20133 Milano, Italy}


\preprint{\textsf{published in Phys.~Rev.~A~{82}, 036101 (2010)}}

\begin{abstract}
\vspace*{.1ex}Inconsistencies are pointed out in a recent proposal [L.~Di\'osi, Phys. Rev. A {\bf 80}, 064104 (2009)]
for a quantum version of
the classical linear Boltzmann equation.
\end{abstract}


\maketitle
\section{Introduction}

In a recent Brief Report {\cite{Diosi2009a}} Di\'osi proposes a quantum
version of the classical linear Boltzmann equation. As explained below, we
think that this proposed equation has a number of unfavorable properties, both
from a physical and from a conceptual point of view, which cast serious doubts
on its validity. This is the more remarkable, as all these problems are
nonexistent in what we consider the appropriate quantum linear Boltzmann
equation (QLBE) {\cite{Hornberger2006b,Hornberger2008a,Vacchini2009a}}, while
the proposed equation has no physically meaningful advantage over the QLBE. In
particular, the possible wish for linearity in the gas momentum distribution,
which serves as the motivation of {\cite{Diosi2009a}}, is consistently
fulfilled within the QLBE by an approximation that is much less drastic than
the modifications entailed in {\cite{Diosi2009a}}.

We are concerned with a Markovian quantum master equation for the motion of a
single, distinguished test particle (mass $M$) due to collisions with a
stationary, homogeneous, and ideal background gas of distinguishable particles
(mass $m$, density $n_{\tmop{gas}}$). The latter is characterized by the
momentum distribution function $\mu_{\tmop{gas}} \left( \tmmathbf{p} \right) =
\langle \tmmathbf{p}| \rho_{\tmop{gas}} |\tmmathbf{p} \rangle \bignone
\bignone$, which may or may not be given by the Maxwell distribution. The
equation is supposed to be valid as long as a Markovian description of the
reduced quantum state of the test particle is appropriate.

\section{Discussion of the master equation}

Di\'osi's proposal involves an ``intercollision time''
\begin{eqnarray}
  \tau & = & \frac{1}{n_{\tmop{gas}} \sigma} \sqrt{\frac{\pi m}{k_{\text{B}}
  T}},  \label{eq:taudef}
\end{eqnarray}
and his master equation is constructed such that in the limit of vanishing gas
density $n_{\tmop{gas}} \rightarrow 0$, i.e. for $\tau \rightarrow \infty$, it
reduces to the classical linear Boltzmann equation for the diagonal momentum
matrix elements. In the following, we do not discuss our reservations about
the derivation of the proposed equation and the implied violation of energy
conservation in individual scattering interactions, but we consider only problems
of the final master equation at finite $\tau$. We start by exposing conceptual
deficiencies and then move on to inconsistencies in the predicted physical
behavior.

\subsection{Conceptual deficiencies}

\subsubsection{Off-shell extension ill-defined}
The equation is not
well-defined because it involves the elastic scattering amplitude $f \left(
\tmmathbf{p}_f, \tmmathbf{p}_i \right)$ at off-shell values $\left|
\tmmathbf{p}_i \right| \neq \left| \tmmathbf{p}_f \right|$. While it is common
in scattering theory to work with off-shell extensions of the scattering
operators, this is only done for computational convenience
{\cite{Taylor1972a}}. All physically relevant properties of the elastic
scattering process depend solely on the on-shell values, not least because of
the arbitrariness of the off-shell extension {\cite{Taylor1972a}}. Even if one
were to agree on a definite choice of the off-shell extension, the proposed
equation would remain ill-defined because the value of the off-shell energy
parameter, which is an independent variable, remains unspecified.

From a physical point of view, it seems implausible that the supposed
``energy-uncertainty'' related to finite intercollision times (a property
involving the state of motion of gas and test particle) has anything to do
with a possible off-shell extension of the elastic scattering amplitude (which
is a function of the interaction potential only).

\subsubsection{Dependence on representation of $\delta$-function}
The
proposed equation depends on a particular choice of the representation of the
$\delta$-function. It has the property that its square can be related to
another representation of the $\delta$-function. One can envisage
representations of the $\delta$-function quite different from the one used in
the proposed equation, which share the same property, e.g. a Gaussian
function. Those would yield manifestly different equations (e.g. leading to a
different prediction for the constant $D_{xx}$). This highlights the
arbitrariness of using a particular representation. From a physical point of
view, it seems implausible that one form of ``smoothing'' should be favored
over another.

\subsubsection{Non-linearity in the gas density}
The proposed equation
is non-linear with respect to the gas density since the definition
(\ref{eq:taudef}) of $\tau$ includes $n_{\tmop{gas}}$. From a physical point
of view, it seems implausible that the Liouvillian for the Markovian dynamics
should be a non-linear function of $n_{\tmop{gas}}$. The reason is that the
background gas is non-interacting and non-degenerate, implying that the gas
particles are uncorrelated, while three-particle collisions are excluded by
assumption. Therefore, each gas particle affects the test particle equally, so
that its effect is described by the same mapping. This implies that the
Liouvillian for the effect of the total gas is proportional to the number of
gas particles, rendering the master equation linear in $n_{\tmop{gas}}$.

\enlargethispage{5ex} 

\subsubsection{Definition of $\tau$}
If the quantity $\tau$ is to
correctly describe the time elapsing between collisions among gas and test
particle, it should not depend only on the state of the gas (via the
temperature in (\ref{eq:taudef}) or more generally via $\mu_{\tmop{gas}}
\left( \tmmathbf{p} \right) = \langle \tmmathbf{p}| \rho_{\tmop{gas}}
|\tmmathbf{p} \rangle$), but it must also depend on the motional state of the
test particle. The definition (\ref{eq:taudef}) is therefore inappropriate, in
particular if the test particle is much faster than the gas particles. In
addition, (\ref{eq:taudef}) is not well-defined because the energy dependence
of $\sigma_{\tmop{tot}}$ is not specified. The physically meaningful mean
intercollision time is given by the expression {\cite{Huang1987}}
\begin{eqnarray}
  \tau_{\tmop{phys}}^{- 1} \left[ \rho \right] & = & \int \mathd \tmmathbf{P}
  \langle \tmmathbf{P}| \rho |\tmmathbf{P} \rangle \, n_{\tmop{gas}} \int \mathd \tmmathbf{p}\,
  \mu_{\tmop{gas}} \left( \tmmathbf{p} \right)  \label{eq:tauprop}\\
  &  & \times v_{\tmop{rel}} \left( \tmmathbf{p}, \tmmathbf{P}
  \right) \sigma_{\tmop{tot}} \left( E_{\tmop{rel}} = {m_{\ast} v_{\tmop{rel}}^2
  \left( \tmmathbf{p}, \tmmathbf{P} \right)/2 }  \right), \nonumber
\end{eqnarray}
with $v_{\tmop{rel}} \left( \tmmathbf{p}, \tmmathbf{P} \right) =
|\tmmathbf{p}/ m -\tmmathbf{P}/ M|$ the relative velocity, and $m_{\ast}$ the
reduced mass. However, the use of the appropriate definition
(\ref{eq:tauprop}) in {\cite{Diosi2009a}} would yield a non-linear time
evolution equation for $\rho$, thus violating the basic requirement of a
linear quantum state evolution.

From a physical point of view, the dependence of the collision rate on the
state of the test particle is rather important, not least because the total
cross section may depend strongly on $E_{\tmop{rel}}$ for large test particle
velocities. Moreover, this dependence is necessary to grant the approach to
the stationary solution. We note that the QLBE discussed in
{\cite{Hornberger2006b,Hornberger2008a,Vacchini2009a}} incorporates this
dependence of the collision rate on $\rho$ (by means of a rate operator with
expectation value $\tau_{\tmop{phys}}^{- 1} \left[ \rho \right]$, see Eq.~(12)
in {\cite{Hornberger2008a}}), while the required linearity in $\rho$ is a
result of the use of concepts from the theory of generalized quantum
measurements {\cite{Hornberger2007b}}.

\subsubsection{Dynamics of momentum populations}
As a natural
consistency requirement, one expects that once the test particle state is
indistinguishable from a classical phase space distribution, its dynamics
should be governed by the classical linear Boltzmann equation (with a quantum
mechanical scattering cross section). This is not fulfilled in the proposed
equation since at finite $n_{\tmop{gas}}$ the dynamics of the populations in
the momentum representation differs from the one predicted by the classical
equation, implying a discontinuous transition to the classical description at
a finite gas density.

\subsubsection{No canonical stationary solution}
As a consequence of Sec.~II A 5, the canonical thermal state of the test particle, $\rho \propto
\exp \left( - \beta \mathsf{P}^2 / 2 M \right)$, is not a stationary solution
of the proposed master equation in a Maxwell-Boltzmann gas. In particular, the
collision kernel does not satisfy the detailed balance condition. From a
physical point of view, it seems quite important that the state of maximal
entropy corresponds to the stationary solution. This is the case even in
extensions of the QLBE which account for quantum degeneracies in the gas
{\cite{Vacchini2009a}}.

\subsubsection{Infinite ``position diffusion'' for $n_{\tmop{gas}} \rightarrow 0$}
The limit of quantum Brownian motion of the proposed master
equation predicts a ``position diffusion'' coefficient $D_{xx}$ (given in Eq.
(23) of {\cite{Diosi2009a}}) that grows above all bounds as one decreases the
gas density $n_{\tmop{gas}} \rightarrow 0$. However, one expects the free
Schr\"odinger equation to be obtained in the limit of vanishing gas density,
so that the predicted behavior is obviously unphysical.

\subsection{Supposed linearity in the gas momentum distribution}
Let us
now comment on the supposed linearity in the gas momentum distribution
discussed in {\cite{Diosi2009a}}. First, we note that (\ref{eq:taudef}) is a
function of the gas temperature and therefore in general a functional of the
gas momentum distribution, \ $\tau = \tau \left[ \mu_{\tmop{gas}} \right]$.
Writing the gas state dependence of the proposed equation in a consistent
fashion, one thus finds that the proposed equation is a manifestly non-linear
expression in $\mu_{\tmop{gas}}$, in stark contrast to the claim in
{\cite{Diosi2009a}}. The advertised advantage of the proposed equation is
therefore unfulfilled.

At the same time, it seems doubtful whether one can or should expect linearity
in $\mu_{\tmop{gas}}$ within a Markovian description. A linear behavior of the
reduced time evolution with respect to an initial, uncorrelated
$\rho_{\tmop{gas}} \left( 0 \right)$ can be expected only based on an exact
solution for system plus environment, while a Markovian description
necessarily involves approximations, required to obtain a Lindblad structure
of the generator, granting complete positivity of the quantum evolution
{\cite{Vacchini2001a}}. The linearity in $\mu_{\tmop{gas}}$ is recovered
easily in the framework of the QLBE once the test particle momentum operator
can be approximated by a characteristic $\mathbbm{C}$-number in the argument
of $\mu_{\tmop{gas}}$.

\enlargethispage{\baselineskip} 

\subsection{Momentum decoherence} We now consider the
  ``apparently overlooked'' ``surprising collisional decoherence
  effect'' advocated in {\cite{Diosi2009a}}.  It is important to
  stress that, far from being overlooked, the effect of the momentum
  exchange due to a single collision 
  is fully accounted for in the QLBE, as
  already discussed in
  {\cite{Vacchini2009a,Breuer2007c,Vacchini2007d}}.  The decoherence
  rate for a hypothetical superposition of two different momentum
  states is predicted by the QLBE to be the average of the
  corresponding total collision rates. This is indeed required on
  physical grounds since any collision changes the momentum by
  definition, and it is immediately seen by looking at the momentum
  representation of the equation, Eq.~(2.7) in
  \cite{Vacchini2009a}:  Considering the off-diagonal element that
  characterizes the coherence, one notes that the ``gain terms'' do
  not affect its temporal change because for an initial superposition
  of momentum eigenstates there are no other off-diagonal elements, so
  that the initial decoherence rate is given by the ``loss terms"
  arising from the anticommutator, which amount to the arithmetic mean
  of the total collision rates. Since there are no stronger physical decoherence mechanisms available, we consider any prediction beyond this as unphysical.

We emphasize that the detailed derivation of the QLBE applies dynamic
scattering theory to a wave packet decomposition of the relative motion
{\cite{Hornberger2008a}}. As such, all the details of the dynamic quantum
scattering process are incorporated by construction, including the particles'
energy uncertainty and the finite interaction time. There is no room for
additional effects in the framework of the two-particle Schr\"odinger
equation. The limits of this treatment are met precisely if the test particle
can never be considered asymptotically free between two collisions because it
interacts with more than one gas particle all the time. The notion of an
intercollision time is meaningless in such situations, and it would be wrong
to use scattering theory altogether.

\subsection{Non-Markovian extensions} As for non-Markovian
  extensions, it is clear that the QLBE loses its validity at
  densities and temperatures where the collisions can no longer be
  taken as independent two-particle events, since the interactions
  between the gas particles start to play a role and the collisions
  cannot be considered as complete. The related problem of the
  derivation of non-Markovian quantum kinetic equations for the
  description of self-interacting dense gases has been 
  considered e.g. in \cite{Kremp1997a,Morozov1999a}, relying on advanced
  many-body techniques. In such frameworks non-Markovian quantum
  extensions of the non-linear Boltzmann equation have been obtained, indicating 
  that memory effects in dense gases
  beyond the Born approximation generally involve nontrivial memory
  kernels.  The difficulties listed above in A.1--A.7 might be avoidable by  
  pursuing an analogous microscopic approach for the QLBE,
  rather than introducing arbitrary ad-hoc modifications.  It thus
  appears that entirely new theoretical approaches are required at high densities, 
  not least since non-Markovian effects or
  corrections cannot possibly be described by a master equation in
  Lindblad form with time independent coefficients
  {\cite{Breuer2009a}}.  In any case, it seems reasonable to use the
properties of the QLBE, including its various limiting forms and its
experimental confirmations discussed in {\cite{Vacchini2009a}}, as a
benchmark consistency check for any proposed extension of that master
equation.

\section{Conclusions} In summary, the equation proposed in
{\cite{Diosi2009a}} provides no physically meaningful advantage over
established results, while displaying conceptual inconsistencies, as
well as leading to unphysical predictions. We conclude that,
rather than bringing ``a dramatic change in our understanding of the quantum
behavior of the test particle'', the equation in {\cite{Diosi2009a}} is a step
in the wrong direction.
\\[\baselineskip]
B.V. acknowledges financial support by Ministero dell'Istruzione, dell'Universit{\`a} e della Ricerca (MIUR), under PRIN2008.

\end{document}